\documentclass[aps,prl,twocolumn]{revtex4}
\usepackage[dvips]{graphicx}
\usepackage{amssymb}
\usepackage{amsmath}
\usepackage{color}

\usepackage[normalem]{ulem}
\usepackage{gensymb}

\usepackage[shortlabels]{enumitem}
\usepackage{tikz}

\begin{document}

\draft

\title{\bf Families of isospectral and isoscattering quantum graphs}

\author{Pavel Kurasov,$^{1}$ Omer Farooq,$^{2}$ Micha{\l} {\L}awniczak,$^{2}$ Szymon Bauch,$^{2}$ Mats-Erik Pistol,$^{3}$ Matthew de Courcy-Ireland,$^{1,4}$ and Leszek Sirko$^{2}$}

\affiliation{$^{1}$Department of Mathematics, Stockholm University, S-106 91 Stockholm, Sweden\\
$^{2}$Institute of Physics, Polish Academy of Sciences, Aleja Lotnik\'{o}w 32/46, 02-668 Warszawa, Poland\\
$^{3}$Solid State Physics, Lund University, S-221 00, Lund, Sweden\\
$^{4}$Nordic Institute for Theoretical Physics, S-106 91 Stockholm, Sweden}

\date{\today}

\begin{abstract}
A concept of germ graphs and the M-function formalism are employed to construct large families of isospectral and isoscattering graphs.  This approach represents a complete departure from the original approach pioneered by Sunada, where isospectral graphs are obtained as quotients of a certain large symmetric graph. Using the M-function formalism and the symmetries of the graph itself we construct  isospectral and isoscattering pairs.  In our novel
approach isospectral pairs do not need to be embedded into a larger	symmetric graph as in Sunada’s approach. We demonstrate that the introduced formalism  can also be extended to graphs with dissipation. The theoretical predictions are validated experimentally using microwave networks emulating open quantum graphs with dissipation.
\end{abstract}


\maketitle
\vspace*{-5,0mm} 

{\it Introduction.-} The concept of quantum graphs was introduced by Pauling \cite{Pauling1936} to describe organic molecules. The great importance of quantum graphs lies in their use as model systems for describing physical networks.
Quantum graphs have been successfully applied to model and study a wide range of physical and mathematical problems and theories, including quantum chaos, non-Weyl graphs, inverse problems, isospectrality and isoscattering, superconductivity theory, microelectronics, photonics crystals, and more \cite{Exner1996,Kottos1997,Kottos2000,Hul2004,Bialous2016,Lawniczak2019, Kostrykin2004,Blumel2002,Pistol,Hul2012,Gnutzmann2006,Berkolaiko2013,Pluhar2013,Chen2021,Wang2023,KuBook,Kowal1990,Imry1996,Rohde2011,Elster2015}. Moreover, the advent of epitaxy techniques has made possible the design and fabrication of quantum nanowire networks, as evidenced by \cite{Samuelson2004,Heo2008}.

The genesis of the isospectrality problem dates back to 1964 and involved Riemannian quotient manifolds \cite{Milnor1964}. This problem has become a huge scientific challenge, when Mark Kac posed a renowned question: ``Can one hear the shape of a drum?" \cite{Kac1966}. The issue was the uniqueness of the Laplace spectrum in the planar domain with Dirichlet boundary conditions. Although  the isospectral Riemannian manifolds were described by Sunada \cite{Sunada1985} in 1985, it took a further seven years to determine how to create pairs of isospectral domains in $\mathbb{R}^2$ \cite{Gordon1992,Sridhar1994,Dhar2003}. Therefore, Kac's question was answered in the negative.

The question of isospectrality in relation to quantum graphs was studied by Gutkin and Smilansky \cite{Gutkin2001} and Kurasov-Nowaczyk \cite{Kurasov2005}. By analyzing the trace formula for quantum graphs, they showed that a graph can be reconstructed from its spectrum if the lengths of its edges are incommensurable. Therefore, in order to obtain non-trivial examples of isospectral graphs, the condition of rational independence of edges should be relaxed.

One of the most powerful approaches to construct isospectral graphs follows Sunada’s idea, where starting with a large graph with non-trivial symmetry group one obtains isospectral graphs as quotient graphs \cite{Parzanchevski2010}. This idea was used by Band, Sawicki, and Smilansky \cite{Band2010,Band2011} to
construct isospectral and isoscattering  quantum graphs with attached infinite leads.
By definition, graphs are isopolar if their scattering matrices have the same poles, or isophasal if the phases of the determinants of their scattering matrices are the same \cite{Band2010,Lawniczak2014}.
Mugnolo and Pivovarchik \cite{Mugnolo2023} show that isospectral graphs on a small number of vertices can be distinguished by attaching a single lead and comparing their scattering properties.	
The existence of topologically distinct isoscattering microwave networks that emulate isoscattering quantum graphs has also been confirmed in a series of experimental tests \cite{Band2010,Hul2012,Lawniczak2014,Lawniczak2021}.	

In this letter, we present a novel approach to isospectrality and isoscattering. By employing the concept of germ graphs and M-function theory \cite{AvKu,KuBook}, we demonstrate how arbitrary graphs, including those with dissipation, can be analyzed to ascertain their isospectrality or isoscattering and to transform them either into the families of isospectral  or  isoscattering graphs or graphs possessing both properties. The germ graphs can exhibit a high degree of simplicity, making the theoretical construction of isoscattering or isospectral graph families explicit.  Therefore, the identification of simple germ networks can even be done experimentally.
\vspace*{-0,6mm}

{\em Quantum graphs.-} Let us consider a metric graph $\Gamma=(V,E)$, which consists of a set $ V = \{v_{m}\}$ of vertices, and 
a set  $ E = \{e_{n}\}$ of edges of the lengths $l_n$. 
Each vertex $v_{i}$ of a graph is connected to the other vertices by $d_i$ edges, where $d_i$ is called the valency of the vertex $v_{i}$.  
 A metric graph becomes quantum when it is equipped  with the Schr\"odinger differential operator  $L(\Gamma)=-\frac{d^2}{dx^2}$ acting on the edges. We will consider the standard (also called Kirchhoff or Neumann \cite{Berkolaiko2013,KuBook}) and Dirichlet vertex conditions. 
 The first one imposes the continuity of waves propagating in edges meeting at $v_{i}$ and the vanishing of the sum of
 their derivatives calculated at the vertex $v_{i}$. The latter requires that the waves vanish at the vertex.
 
 Questions of isospectrality and isoscattering have been studied for other differential operators such as Dirac operators \cite{Harrison2024} and buckling operators from solid state physics \cite{Kurasov2021}, not only for metric graphs but also for domains in Euclidean space and surfaces \cite{Zworski2009}. Our approach is based on symmetries and therefore can be applied in these settings as well.

\begin{figure}[h]
\includegraphics[width=1.0\linewidth]{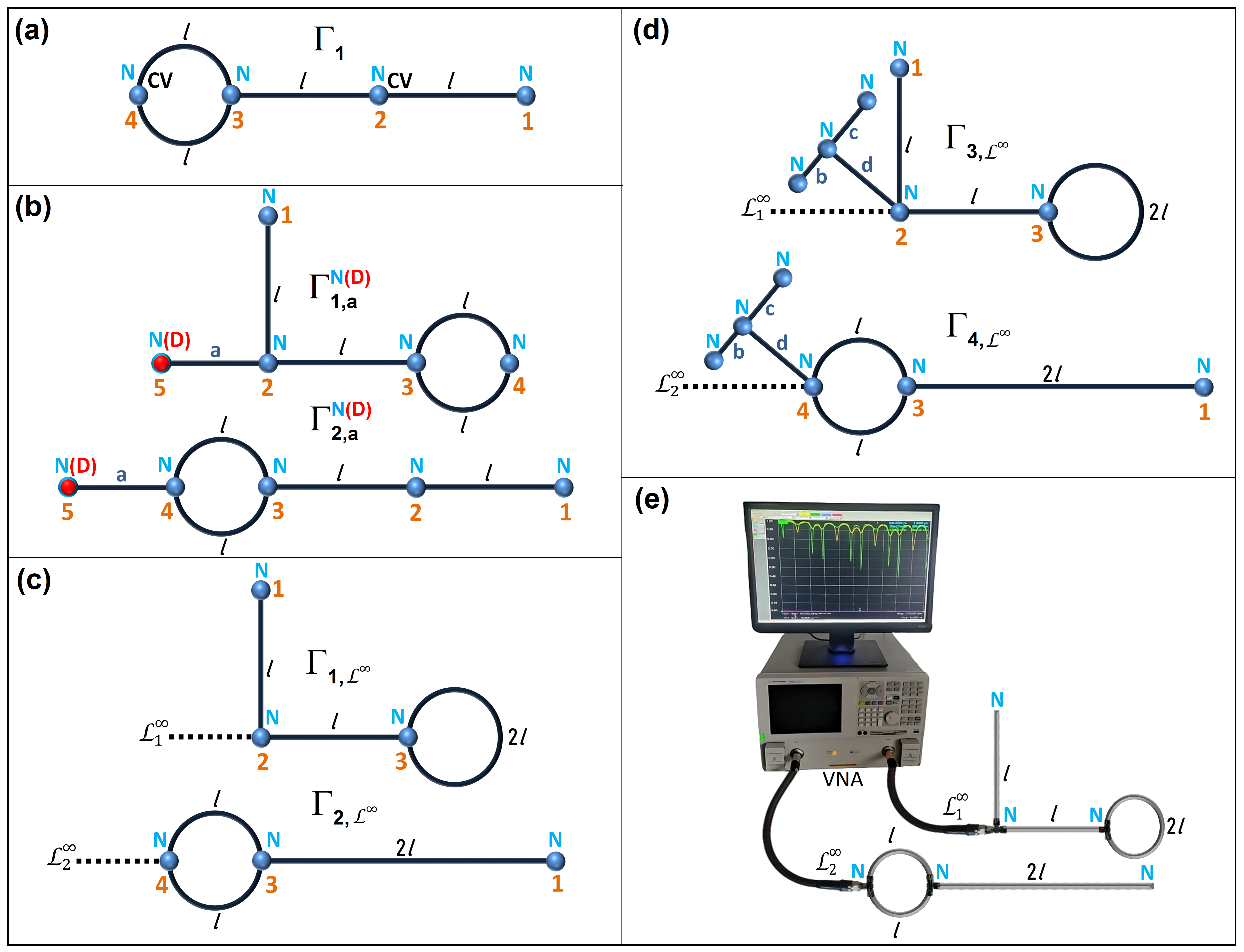}
\caption{
(a) A germ graph $\Gamma_1$ with Neumann boundary conditions (blue capital letters N) at the vertices. Contact vertices $v_2$ and $v_4$ are marked by CV. (b) Two pairs of graphs $\Gamma_{\rm 1,a}^{\rm N}$ and $\Gamma_{\rm 2,a}^{\rm N}$ with Neumann and $\Gamma_{\rm 1,a}^{\rm D}$ and $\Gamma_{\rm 2,a}^{\rm D}$ with Dirichlet boundary conditions (red capital letters D) at the vertices No. 5.  (c) Schemes of isoscattering microwave networks $\Gamma_{1,\mathcal{L}^{\infty}}$ and $\Gamma_{2,\mathcal{L}^{\infty}}$ with Neumann  boundary conditions.  The networks are connected to a microwave vector network analyzer via leads $\mathcal{L}^{\infty}_1$ and  $\mathcal{L}^{\infty}_2$, respectively. (d) Schemes of isoscattering microwave networks $\Gamma_{3,\mathcal{L}^{\infty}}$ and $\Gamma_{4,\mathcal{L}^{\infty}}$ with Neumann  boundary conditions. They are the members of isoscattering family built on the germ graph $\Gamma_1$. A three edge star graph is connected to the contact vertices of the graph $\Gamma_1$. (e) Microwave networks $\Gamma_{1,\mathcal{L}^{\infty}}$ and $\Gamma_{2,\mathcal{L}^{\infty}}$ with Neumann boundary conditions.  Both networks are characterized by the same total length $L=4l=1.1580 \pm 0.0012$ m. The networks are connected to a microwave vector network analyzer marked by VNA via microwave cables (leads) $\mathcal{L}^{\infty}_1$ and  $\mathcal{L}^{\infty}_2$, respectively.}
\label{Fig1}
\end{figure}

{\it Scattering on compact graphs.-} With any metric graph $ \Gamma $ and any finite set of graph vertices $\partial \Gamma$, called contact vertices, we may associate a scattering matrix $S_{\Gamma}(k)$ in the following sense (see Fig. \ref{Fig1}(a)).
Let us construct another (non-compact) graph $ \Gamma^{\rm ext} $ obtained by attaching to $ \Gamma $
precisely
$ | \partial \Gamma | $ infinite leads - the semi-infinite intervals parametrised as $ [0, \infty) $.
We assume that standard vertex conditions are introduced in the new vertices
obtained by adding left end-points of the semi-infinite intervals to each of the contact vertices from $ \partial \Gamma $.
The corresponding stationary scattering matrix $S_{\Gamma}(k)$ connects the amplitudes of the incoming $ \vec{a} $ and outgoing $ \vec{b} $ plane waves:
$ \vec{b} =	S_{\Gamma}(k) \vec{a} $,
where the vectors $ \vec{a}$ and $\vec{b} $ have dimension $ |\partial \Gamma | $.

{\it M-functions.~-} Titchmarsh-Weyl M-functions are a powerful tool in the spectral theory of one-dimensional Schr\"odinger operators \cite[Ch. 17]{KuBook}. Their application to metric graphs allows to describe spectra of graphs and to solve inverse problems \cite{Pistol}.

We use the following formula to define the M-function associated with the contact  vertices $ \partial \Gamma$
\begin{equation}
M_\Gamma (k) : \vec{u}_{\partial \Gamma} \mapsto  \partial \vec{u}_{\partial \Gamma},
\end{equation}
where  $ \vec{u}_{\partial \Gamma}, \partial \vec{u}_{\partial \Gamma} $ denote the values of the function and the sums of normal derivatives at the vertices from the contact set $ \partial \Gamma$ for any function $u$ satisfying the eigenfunction equation with spectral parameter $k^2$ inside the graph and having the prescribed values $\vec{u}_{\partial \Gamma}$ at the contact vertices.
If the contact set consists of a single vertex $v$, the definition can be simplified:
$M_\Gamma(k) = \partial u(v)/u(v)$,
where $u(v)$ and $\partial u(v)$ denote the values of the function and the sum of the normal derivatives
at the contact vertex $v$.
 
The M-function formalism provides a straightforward methodology for  analyzing the process of graph gluing.   Let us consider a graph $\Gamma$ which is constructed by gluing  two metric graphs $\Gamma_1$ and $\Gamma_2$ at the contact vertices $|\partial \Gamma_1|=|\partial \Gamma_2|$. The M-function for the glued graph is equal to the sum of the M-functions associated with the original parts:
\begin{equation}
M_{\Gamma}=M_{\Gamma_1}+M_{\Gamma_2}.
\label{SumM}
\end{equation}

For any metric graph $\Gamma$, the S-matrix and the M-function are related as follows (formula ({18.40}) in \cite{KuBook})
\begin{equation} 
	 S_\Gamma (k) = \frac{ ikI - M_\Gamma (k)}{ikI + M_\Gamma (k)}, 
	 \label{S}
\end{equation}
where $I$ is the  identity matrix.
 This formula determines $ | \partial \Gamma | \times | \partial \Gamma | $
matrix valued function. Since for any  dissipation-free metric graph $\Gamma$,  $M_\Gamma (k)$ is Hermitian almost everywhere, $ S_\Gamma (k) $ is unitary. 

The M-function is related to the Wigner reaction matrix $K$ \cite{Fyodorov2004,Savin2005}, often used in the theory of nuclear reactions. This connection is especially straightforward in the case of quantum graphs, $M=-kK$, where $M$ and $K$ are matrix-valued functions.

	
{\it Isoscattering and isospectrality.-}
The M-function formalism allows to tell when a pair of graphs are isoscattering (or isospectral), and enables the general construction of families of such graphs. 
In this article we will consider the most commonly used definition of isoscattering graphs.
Two metric graphs $\Gamma_1 $ and $\Gamma_2 $ with contact sets $\partial \Gamma_1 $ and
$\partial \Gamma_2 $ are isoscattering if
	$ S_{\Gamma_2} (k) =   S_{\Gamma_1}(k)$.
Isoscattering implies in particular that the graphs have the same number of
contact vertices
$ | \partial \Gamma_1 | = | \partial \Gamma_2 |. $

Formula \eqref{S} implies that two graphs  $\Gamma_1 $ and $\Gamma_2 $ are isoscattering if  
 	$M_{\Gamma_2} (k) =   M_{\Gamma_1}(k)$.
If this relation holds, graphs $\Gamma_1$ and $\Gamma_2$ are isoscattering, even if they are composed of any number of compound graphs.
 
Two metric graphs $\Gamma_1 $ and $\Gamma_2 $ are isospectral if they have the same spectra, counting multiciplicities. Isoscattering graphs are not necessarily isospectral.  The spectra may differ because of eigenfunctions that satisfy both Dirichlet and standard conditions at the contact vertices. Such eigenfunctions and their associated eigenvalues are invisible to the M-function.
To ensure that two isoscattering graphs are isospectral one should in addition require that
the invisible spectra coincide.

{\bf Theorem 1} 
Let two graphs $\Gamma_1$ and  $\Gamma_2$ be isoscattering  and let $ \Gamma $ be any other graph
with the same number of contact vertices $|\partial \Gamma|=|\partial \Gamma_1|=|\partial \Gamma_2|$.
Consider the glued graphs  $\Gamma_1^f=\Gamma_1 \sqcup \Gamma$ and $\Gamma_2^f=\Gamma_2 \sqcup \Gamma$,
obtained by pairwise identifying contact vertices in $ \partial \Gamma_j $ and $ \partial \Gamma $. Then it holds:
\vspace*{-6,5mm}
\begin{enumerate}[(a)]
	\item 
	The graphs $ \Gamma_1^f $ and $ \Gamma_2^f $, with contact sets given by the glued vertices, are isoscattering.
	Their visible spectra coincide and are defined by the secular equations
	\vspace*{-2,0mm}
	\begin{equation}
		\det (M_{\Gamma_j}(k) + M_{\Gamma}(k) )= 0.
	\end{equation}	
\vspace*{-9,0mm}
	\item
	 The graphs 
	$ \Gamma_1^f $ and $ \Gamma_2^f $ are also isospectral if in addition the graphs $ \Gamma_1 $ and $ \Gamma_2 $ were isospectral.
	The invisible eigenvalues are inherited from the original graphs $ \Gamma_j $ and $ \Gamma $.
\end{enumerate}
\vspace*{-2,0mm}
 

{\it Explicit examples.-} We think of $\Gamma_1$ and $\Gamma_2$ as a germ from which families of isospectral or isoscattering graphs can be grown.
As a special case, one can take $\Gamma_1$ and $\Gamma_2$ to be a single graph with two distinct contact vertices having the same M-functions.


{\it Families of isospectral graphs.-} Consider the example of the germ graph presented in Fig.~\ref{Fig1}(a). It contains four vertices with Neumann boundary conditions including two contact vertices $v_2$ and $v_4$, which are additionally marked as $\rm CV$.
According to {\bf Theorem 1(b)} our first step is to derive the M-functions associated with the contact vertices  $ v_2 $ and $v_4$ on the germ  graph $\Gamma_{\rm 1}$.  This involves two more M-functions $M_{\Gamma_{5}}(k)$ and $M_{\Gamma_{6}}(k)$ calculated in the Supplementary Material, formulas (S1) and (S2). 
Equality of the M-functions $M_{\Gamma_{5}}(k) = M_{\Gamma_{6}}(k)$ implies that the contact vertices are isoscattering and the family of the isospectral graphs can be obtained by attaching to the contact vertices any compact graph. 
The realization of simple isospectral graphs $\Gamma_{\rm 1,a}^{\rm N(D)}$ and $\Gamma_{\rm 2,a}^{\rm N(D)}$ are shown in Fig.~\ref{Fig1}(b). They are obtained by attaching to the contact vertices $v_2$ and $v_4$ the interval  of length $a>0$, with one endpoint forming the contact vertex and the Neumann or Dirichlet boundary condition at the
opposite endpoint.
 The M-function for an interval of length $a$ with either Neumann or Dirichlet boundary conditions at the free end is  given by $M_{{\rm a}^{\rm N}} (k) = k \tan ka$ and $M_{{\rm a}^{\rm D}} (k)= -k \cot ka$, respectively. 
Then, the spectra of graphs $\Gamma_{\rm 1,a}^{\rm N(D)}$  and $\Gamma_{\rm 2,a}^{\rm N(D)}$ corresponding to visible eigenfunctions coincide respectively with the zeroes of the following secular equations:
\begin{equation} \label{sec1}
	\begin{array}{ll}
 M_{\Gamma_{\rm 1,a}^{\rm N(D)}}(k) = M_{\Gamma_{5}}(k) +  M_{{\rm a}^{\rm N(D)}} (k) = 0, \\
M_{\Gamma_{\rm 2,a}^{\rm N(D)}}(k) = M_{\Gamma_{6}}(k) + M_{{\rm a}^{\rm N(D)}} (k) = 0. 
\end{array}
\end{equation}
Due to equality of the M-functions $M_{\Gamma_{5}}(k)$ and $M_{\Gamma_{6}}(k)$,  the pairs of graphs $\Gamma_{\rm 1,a}^{\rm N}$  and $\Gamma_{\rm 2,a}^{\rm N}$, and $\Gamma_{\rm 1,a}^{\rm D}$  and $\Gamma_{\rm 2,a}^{\rm D}$ are isospectral.


{\it Families of isoscattering graphs.-} Isoscattering of the contact vertices $v_2$ and $v_4$ in the germ  graph $\Gamma_{\rm 1}$ can be also verified experimentally.  The schemes of isoscattering microwave networks $\Gamma_{1,\mathcal{L}^{\infty}}$ and $\Gamma_{2,\mathcal{L}^{\infty}}$ with Neumann boundary conditions are shown in Fig.~\ref{Fig1}(c). The leads  $\mathcal{L}^{\infty}_1$ and  $\mathcal{L}^{\infty}_2$ in the microwave networks are connected in the contact vertices $v_2$ and $v_4$ defined in Fig.~\ref{Fig1}(a) and in Fig.~1 in the Supplementary Material. The newly created vertices are characterized by Neumann boundary conditions and possess valency $n_c=3$. 

The properties of the germ graph  $\Gamma_{\rm 1}$, as described in {\bf Theorem 1(a)}, permit the construction of a family of isoscattering graphs of arbitrary complexity. Any compact or non-compact graph can be connected to the contact vertices $v_2$ and $v_4$.  
To illustrate such properties, Fig.~\ref{Fig1}(d) presents the schemes of the isoscattering networks  $\Gamma_{3,\mathcal{L}^{\infty}}$  and $\Gamma_{4,\mathcal{L}^{\infty}}$ constructed by attaching to the contact vertices a simple compact star graph $\Gamma_{\rm S}$ (see Fig.~2 in the Appendix) with the edges $b$, $c$, and $d$ which are not in a rational relation with the edges forming the germ graph $\Gamma_1$. The experimental confirmation of their isoscattering is presented below.	
It should also be noted that according to {\bf Theorem 1(a)} the graphs $\Gamma_{3,\mathcal{L}^{\infty}}$  and $\Gamma_{4,\mathcal{L}^{\infty}}$ would be isoscattering for any boundary conditions at the pendant edges $b$ and $c$.
From {\bf Theorem 1(b)} we also find that if in the graphs $\Gamma_{\rm 1,a}^{\rm N(D)}$ and $\Gamma_{\rm 2,a}^{\rm N(D)}$ the edge $a$ is replaced by the star graph $\Gamma_{\rm S}$, the newly formed graphs will be isospectral for any boundary conditions at the pendant edges $b$ and $c$ of the star graph $\Gamma_{\rm S}$.


{\it Experimental observations.-} It has been proved by Hul et al.\ \cite{Hul2004} that
the telegrapher's equation for microwave networks is formally
equivalent to the one-dimensional Schr\"odinger equation describing quantum graphs. That is why quantum graphs can be emulated by microwave networks and the properties of quantum graphs can be studied experimentally using
microwave networks with the same topology and boundary conditions at the
vertices. Similarly to a finite compact quantum graph a microwave network consists of $V$ vertices (microwave joints) connected by $E$ edges (microwave cables) \cite{Hul2004,Lawniczak2019}. 
Microwave networks enable the emulation of a variety of quantum chaotic systems whose spectral properties can be described by the three main symmetry classes: Gaussian orthogonal ensemble (GOE) \cite{Hul2004,Hul2009,Dietz2017,Lawniczak2020,Chen2020}, Gaussian unitary ensemble (GUE) \cite{Hul2004,Allgaier2014,Bialous2016,Lawniczak2019b,Lawniczak2023} and Gaussian symplectic ensemble (GSE) \cite{Rehemanjang2016,Lu2020,Lawniczak2023} within the framework of the Random Matrix Theory.  Currently, the properties of GSE systems can only be studied experimentally using microwave networks.
 
\begin{figure}[h]
	\begin{center}
	\includegraphics[width=0.9\linewidth]{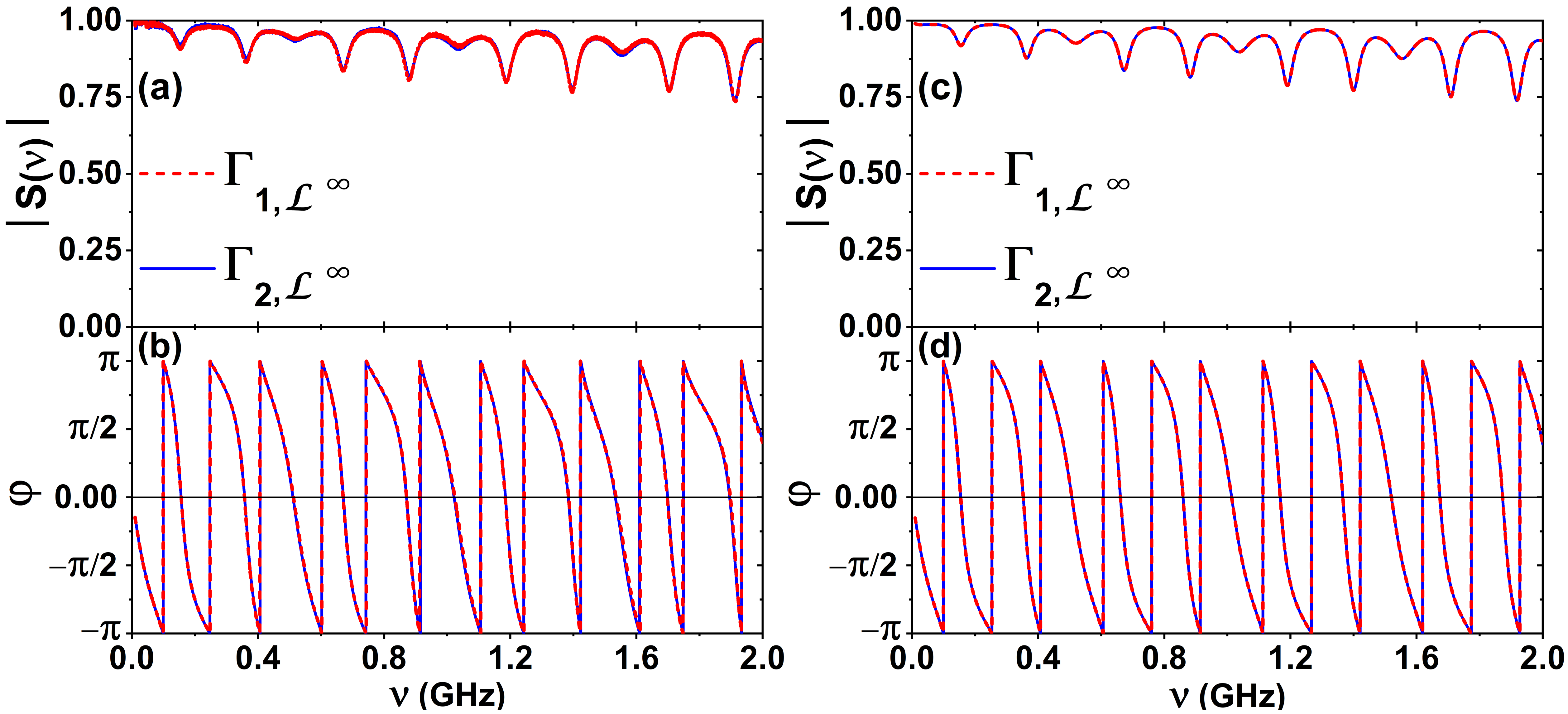}
	\caption{
Panels (a) and (b) show the amplitudes  $|S(\nu)|$ and the phases $\varphi$ of the scattering matrices  $S(\nu)$  of the experimentally studied microwave networks $\Gamma_{1,\mathcal{L}^{\infty}}$  (red broken line) and $\Gamma_{2,\mathcal{L}^{\infty}}$  (blue solid line) in the frequency range 0.01-2 GHz.
(c) The absolute value $|S(\nu)|$ and (d) the phase $\varphi$ of the scattering matrices $S(\nu)$ (see Eq.~(\ref{S})) calculated  for the quantum graphs with dissipation emulating microwave networks  $\Gamma_{1,\mathcal{L}^{\infty}}$ (red broken line) and $\Gamma_{2,\mathcal{L}^{\infty}}$  (blue solid line) in the frequency range 0.01-2 GHz.}
	\label{Fig2}
\end{center}
\end{figure}

\begin{figure}[h]
	\includegraphics[width=0.9\linewidth]{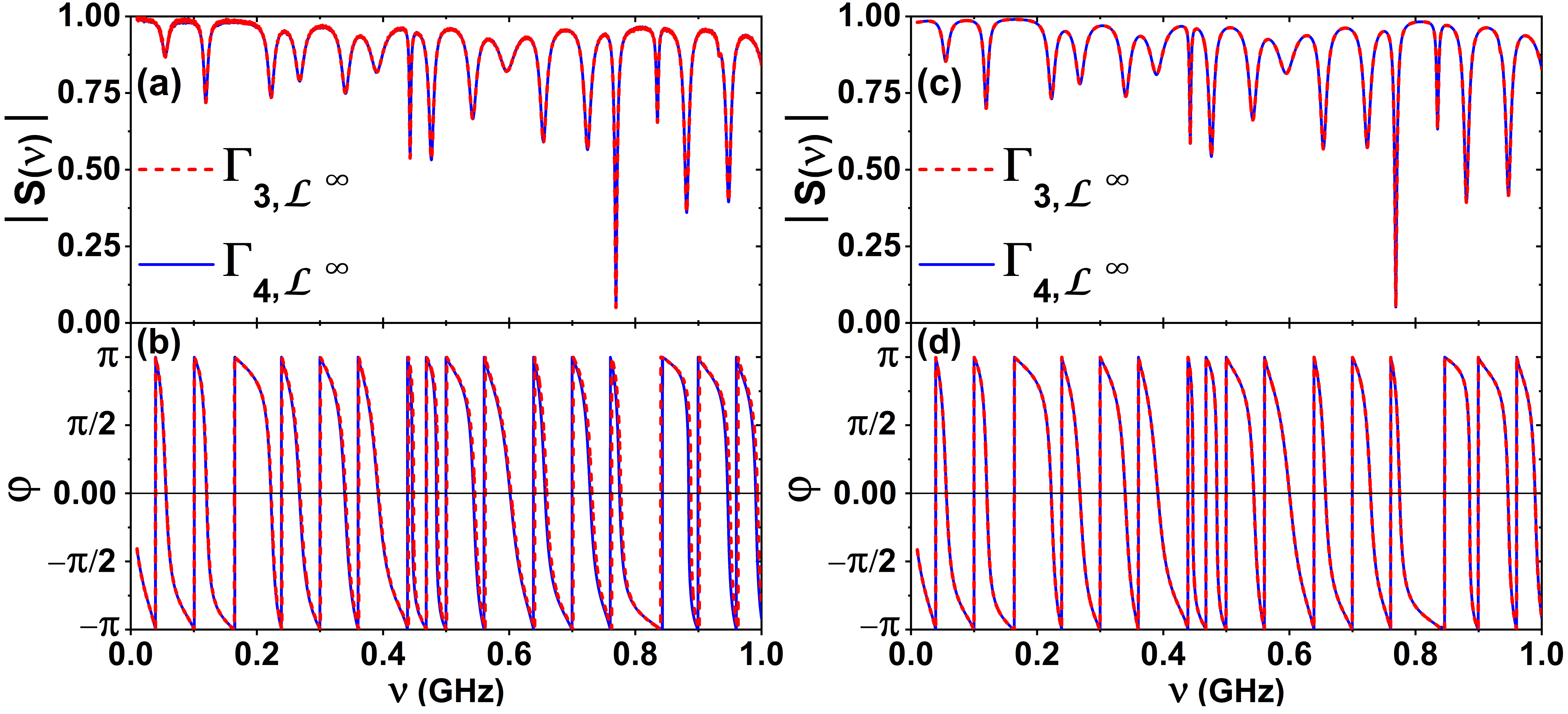}
	\caption{
The amplitudes  $|S(\nu)|$ (panel (a)) and the phases $\varphi$ (panel (b)) of the scattering matrices  $S(\nu)$ of the experimentally studied microwave networks  $\Gamma_{3,\mathcal{L}^{\infty}}$ (red broken line) and  $\Gamma_{4,\mathcal{L}^{\infty}}$ (blue solid line) with Neumann boundary conditions in the frequency range 0.01-1 GHz.
(c) The absolute value $|S(\nu)|$ and (d) the phase $\varphi$ of the scattering matrices $S(\nu)$  calculated  for the quantum graphs with dissipation emulating microwave networks   $\Gamma_{3,\mathcal{L}^{\infty}}$ (red broken line) and  $\Gamma_{4,\mathcal{L}^{\infty}}$ (blue solid line)  with standard boundary conditions in the frequency range 0.01-1 GHz.}
\label{Fig3}
\end{figure}

Microwave realizations of isoscattering graphs $\Gamma_{1,\mathcal{L}^{\infty}}$ and $\Gamma_{2,\mathcal{L}^{\infty}}$ with Neumann boundary conditions are shown in Fig.~\ref{Fig1}(e). The leads  $\mathcal{L}^{\infty}_1$ and  $\mathcal{L}^{\infty}_2$ in the microwave networks are connected in the contact vertices defined in Fig.~\ref{Fig1}(a) and Fig.~1 in the Supplementary Material. The new vertices created in this way are characterized by Neumann boundary conditions and have valency $n_c=3$. 	In order to obtain the one-port scattering matrix $ S(\nu)=|S(\nu)|\exp{(i\varphi)}$, where $|S(\nu)|$ and $\varphi$ are the amplitude and the phase of the scattering matrix, the scattering matrix  $ S(\nu)$ was measured
in the frequency range $\nu = 0.01-2$ GHz (see Fig.~\ref{Fig1}(e) and Fig.~\ref{Fig2}). The wave vector $k$ expressed in terms of frequency $\nu$ is defined as $k=\frac{2\pi \nu}{c}$. Here, $c$ is the speed of light in vacuum. The
connection of a vector network analyzer (VNA) to a microwave network (see Fig.~\ref{Fig1}(e)) is equivalent to attaching
the infinite leads $\mathcal{L}^{\infty}_1$ and $\mathcal{L}^{\infty}_2$, $a \rightarrow \infty$, to the quantum graphs $\Gamma_{\rm 1,a}^{\rm N}$ and $\Gamma_{\rm 2,a}^{\rm N}$, respectively.
In Fig.~\ref{Fig2}(a)-(b) we  show the amplitudes  $|S(\nu)|$ and the phases $\varphi$ of the scattering matrices  $S(\nu)$  of the experimentally studied microwave networks $\Gamma_1^{\rm N}$ (red broken line) and $\Gamma_2^{\rm N}$ (blue solid line) with Neumann boundary conditions in the frequency range 0.01-2 GHz. Our experimental results clearly demonstrate that both graphs are isoscattering.
Panels (a) and (b) in Fig.~\ref{Fig3} show the amplitudes  $|S(\nu)|$ and the phases $\varphi$ of the scattering matrices  $S(\nu)$  of the experimentally studied microwave networks $\Gamma_{3,\mathcal{L}^{\infty}}$ (red broken line) and $\Gamma_{4,\mathcal{L}^{\infty}}$  (blue solid line) with Neumann boundary conditions in the frequency range 0.01-1 GHz.  The microwave networks $\Gamma_{3,\mathcal{L}^{\infty}}$  and $\Gamma_{4,\mathcal{L}^{\infty}}$ are constructed by attaching to the contact vertices of the network $\Gamma_1$ with the total length $L=4l=3.0000\pm 0.0012$ m a simple compact star microwave network $\Gamma_{\rm S}$ (see Fig.~2 in the Supplementary Material) with the edges $b=0.0809 \pm 0.0003$ m, $c=0.2400\pm 0.0003$ m, and $d=0.1909 \pm 0.0003$ m.
The experimental results unambiguously confirm that both networks are isoscattering, as postulated by {\bf Theorem~1(a)}.

In order to theoretically describe the properties of actual microwave networks with dissipation, it is necessary to apply the formalism of dissipative M-functions. This may be accomplished by employing the following substitution of the wave vector \cite{Lawniczak2014} in the formula (\ref{S}), 
 $k \mapsto k + i \beta \sqrt{k}$,  $\quad 0 < \beta \ll 1.$
As a consequence the scattering matrix (\ref{S}) is not unitary (unimodular) anymore.
Its absolute value diverges from $1$ as $ k $ increases with substantial differences near the eigenvalues of the operator on the compact graph.

The corresponding plots can be seen in Fig.~\ref{Fig2}(c)-(d) and Fig.~\ref{Fig3}(c)-(d). In the numerical calculations an absorption coefficient $\beta=0.0055$ m$^{-1/2}$ and $\beta=0.0084$ m$^{-1/2}$ was, respectively, applied.
In  Fig.~\ref{Fig2}(c)-(d) we show  the absolute value $|S(\nu)|$ and the phase $\varphi$ of the scattering matrices $S(\nu)=S_{\Gamma}(\frac{ck}{2\pi})$ (see Eq. (\ref{S})) calculated  for quantum graphs with dissipation emulating microwave networks  $\Gamma_{1,\mathcal{L}^{\infty}}$  (red broken line) and  $\Gamma_{2,\mathcal{L}^{\infty}}$ (blue solid line)  with Neumann boundary conditions in the frequency range 0.01-2 GHz. In the calculations dissipative M-functions  $M_{\Gamma_5}(k) = M_{\Gamma_6}(k)$ were applied. Similarity between the experiment, Fig.~\ref{Fig2}(a)-(b), and theory, Fig.~\ref{Fig2}(c)-(d), is striking! 
In Fig.~\ref{Fig3}(c)-(d)  the absolute value $|S(\nu)|$ and the phase $\varphi$ of the scattering matrices $S(\nu)$ calculated  for the quantum graphs with dissipation emulating microwave networks $\Gamma_{3,\mathcal{L}^{\infty}}$ (red broken line) and $\Gamma_{4,\mathcal{L}^{\infty}}$ (blue solid line)  with Neumann boundary conditions are shown in the frequency range 0.01-1 GHz. In the calculations dissipative M-functions  $M_{\Gamma_5}(k)+M_{\Gamma_{\rm S}}$ and $M_{\Gamma_6}(k)+M_{\Gamma_{\rm S}}$ were applied, respectively. The agreement between the experiment (see Fig.~\ref{Fig3}(a)-(b)) and theoretical results presented in  Fig.~\ref{Fig3}(c)-(d) is very good.

The reported measurements not only verify our theoretical results, but are also indispensable for the expected applications of isospectral or isoscattering networks and the experimental identification of isoscattering germ networks. In such cases, not only dissipation but also the appropriate design of the boundary conditions at the vertices should be taken into account. The agreement with theory means that in some cases the germ networks, and hence the families of complex isoscattering networks, can be constructed solely on the basis of measurements.

{\it Summary.-} We introduced the concept of germ graphs and employed the M-function formalism to derive the necessary conditions for the identification and construction of general families of isospectral and isoscattering quantum graphs. Our analysis concerns both ideal, non-dissipative, and open, dissipative quantum graphs. In the latter case, the theoretical predictions are validated experimentally using microwave networks. 


\smallskip

\end{document}


\draft

\title{\bf SUPPLEMENTARY MATERIAL for\\
	Families of isospectral and isoscattering quantum graphs}

\author{Pavel Kurasov,$^{1}$ Omer Farooq,$^{2}$ Micha{\l} {\L}awniczak,$^{2}$ Szymon Bauch,$^{2}$ Mats-Erik Pistol,$^{3}$ Matthew de Courcy-Ireland,$^{1,4}$ and Leszek Sirko$^{2}$}

\affiliation{$^{1}$Department of Mathematics, Stockholm University, S-106 91 Stockholm, Sweden\\
	$^{2}$Institute of Physics, Polish Academy of Sciences, Aleja Lotnik\'{o}w 32/46, 02-668 Warszawa, Poland\\
	$^{3}$Solid State Physics, Lund University, S-221 00, Lund, Sweden\\
    $^{4}$Nordic Institute for Theoretical Physics, S-106 91 Stockholm, Sweden}

\date{\today}
	
	
	\smallskip

\maketitle

We provide some additional details for the calculations described in the text of the Letter.
\vspace{2mm}
\begin{center}
{\bf Analytic calculations of M-functions}
\end{center}

\subsection{The germ graph $\Gamma_1$ with Neumann boundary conditions}

\begin{figure}[h]
\begin{center}
	\includegraphics[width=1.0\linewidth]{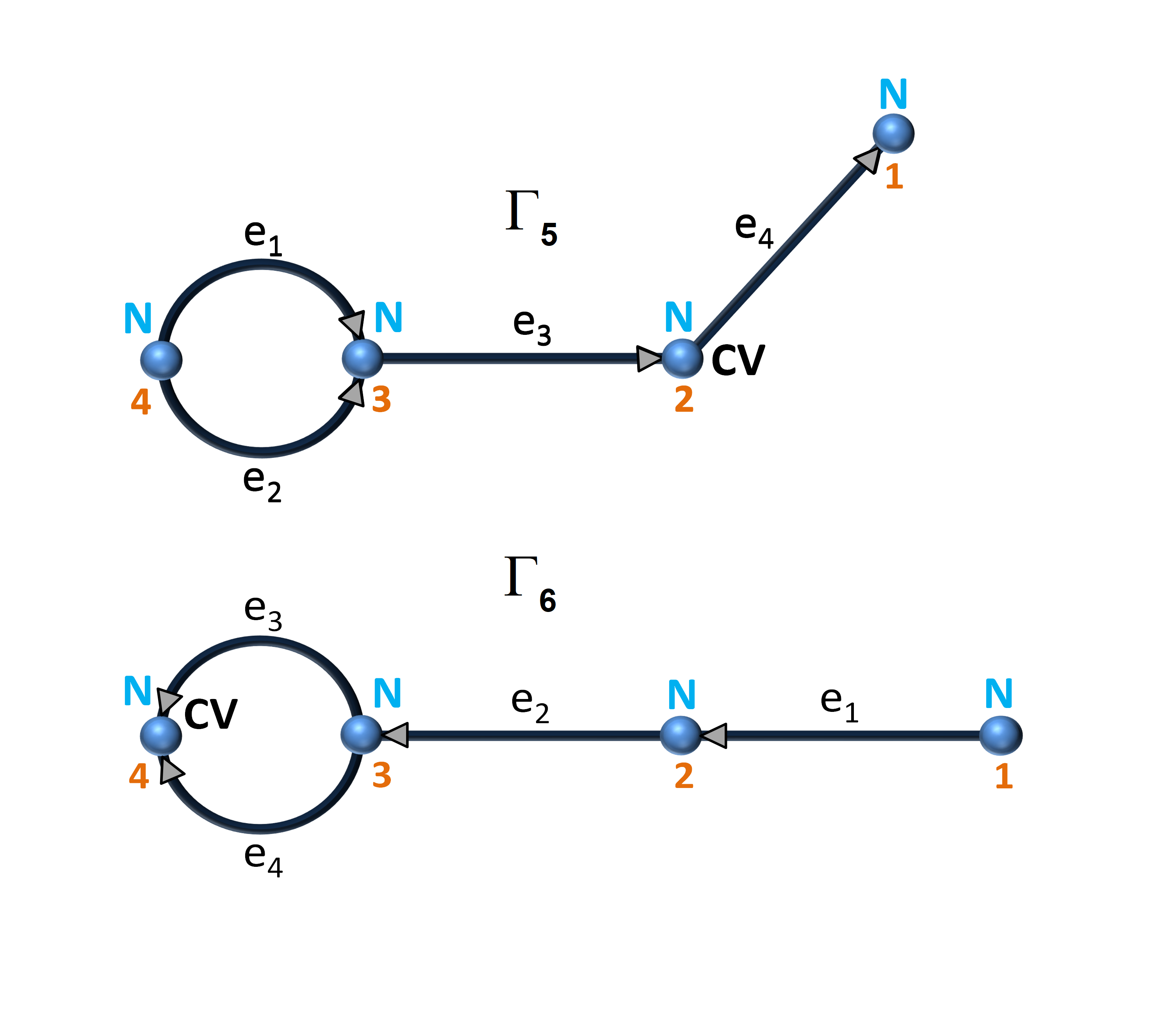}
	\caption{
		A pair of equivalent, isospectral graphs $ \Gamma_5$ and $ \Gamma_6$ with Neumann  boundary conditions and equal M-functions at the contact vertices $v_2$ and $v_4$, respectively,  marked by CV. The vertices with Neumann boundary conditions are denoted by blue capital letters N. All edges $e_i$ have length $\ell$. }
\label{FigS1}
\end{center}
\end{figure}

In order to facilitate a more intuitive understanding of our calculations,  we present in Fig.~\ref{FigS1} the germ graph, denoted in the Letter as $\Gamma_1$ in Fig.~1(a), in the form of two equivalent graphs  $ \Gamma_5$ and  $ \Gamma_6$.

The first graph $ \Gamma_5$ in Fig.~\ref{FigS1} is divided by the contact vertex into two, hence the M-function is equal to the sum of M-functions
for the two components. 

Parametrization of the intervals is indicated by arrows so that the starting end point corresponds to
$ x= 0 $ and the arrow corresponds to $ x=\ell.$
The numbers of the edges are indicated as well, which allows us to denote the
functions on the interval $ e_j $ by $ u^j $.

Calculating the M-function for the left component it is enough
to consider functions symmetric with respect to horizontal line. Antisymmetric functions do not
contribute to the M-function and are therefore invisible.

The following set of functions satisfy the differential equation on the left component formed by $ e_1, e_2 $ and $e_3$, and
Neumann vertex conditions at the inner vertices:
$$
\left\{
\begin{array}{l}
	u^1 =  \cos kx, \\
	u^2  =  \cos kx \\
u^3  =  \displaystyle  \cos( k\ell)  \cos kx -  2 k \sin (k\ell) \frac{\sin kx}{k}  = \\ \cos (k\ell) \cos kx - 2 \sin (k\ell) \sin kx.
\end{array}
\right.
$$
Hence the M-function for the left component is
$$ k  \frac{ \cos (k\ell) \sin (k\ell) +  2 \sin (k\ell) \cos (k\ell)
}{ \cos (k\ell) \cos (k\ell) - 2 \sin (k\ell) \sin (k\ell)
},
$$
where we used that $ \partial u^3 (\ell) = - (u^3)' (\ell). $  Adding the M-function for the interval of length $\ell$ with the Neumann vertex condition,
 $M_{{\ell}^{\rm N}} (k) = k \tan k\ell$, we get
\begin{equation}
		\begin{array}{l}
M_{\Gamma_5} (k) =  k\frac{ \cos (k\ell) \sin (k\ell) +  2 \sin (k\ell) \cos (k\ell)
}{ \cos (k\ell) \cos (k\ell) - 2 \sin (k\ell) \sin (k\ell) 
} + k \tan k\ell = \\
\\
2 k\frac{(2-3\sin^2{k\ell})\sin{k\ell}}{(1-3\sin^2{k\ell})\cos{k\ell}}.
	\end{array}
\tag{S1}
\label{S1}
\end{equation}

In case of the graph $ \Gamma_6$ we again consider only functions that are symmetric with respect to the horizontal line:
$$
\left\{
\begin{array}{l}
	u^1  =  \cos kx, \\
	u^2  =  \cos k(x+\ell) \\
	u^3 = u^4  =  \displaystyle  \cos (2 k\ell)  \cos kx -  \frac{1}{2} k \sin (2 k\ell) \frac{\sin kx}{k}  = \\
	\cos(2  k\ell) \cos kx - \frac{1}{2} \sin( 2 k\ell) \sin kx.
\end{array}
\right.
$$
The corresponding M-function is
\begin{equation}
	\begin{array}{l}
	M_{\Gamma_6} (k) = 2 k \frac{2  \cos2  (k\ell) \sin (k\ell) + \sin 2 (k\ell) \cos (k\ell)}{2  \cos2  (k\ell) \cos (k\ell) - \sin 2 (k\ell) \sin (k\ell)} = \\
	\\
	 2 k\frac{(2-3\sin^2{k\ell})\sin{k\ell}}{(1-3\sin^2{k\ell})\cos{k\ell}}.
	\end{array}
\tag{S2}
\label{S2}
\end{equation}

Thus we have shown that the two graphs are isoscattering.

\subsection{A star graph}

\begin{figure}[h]
	\begin{center}
		\includegraphics[width=1.0\linewidth]{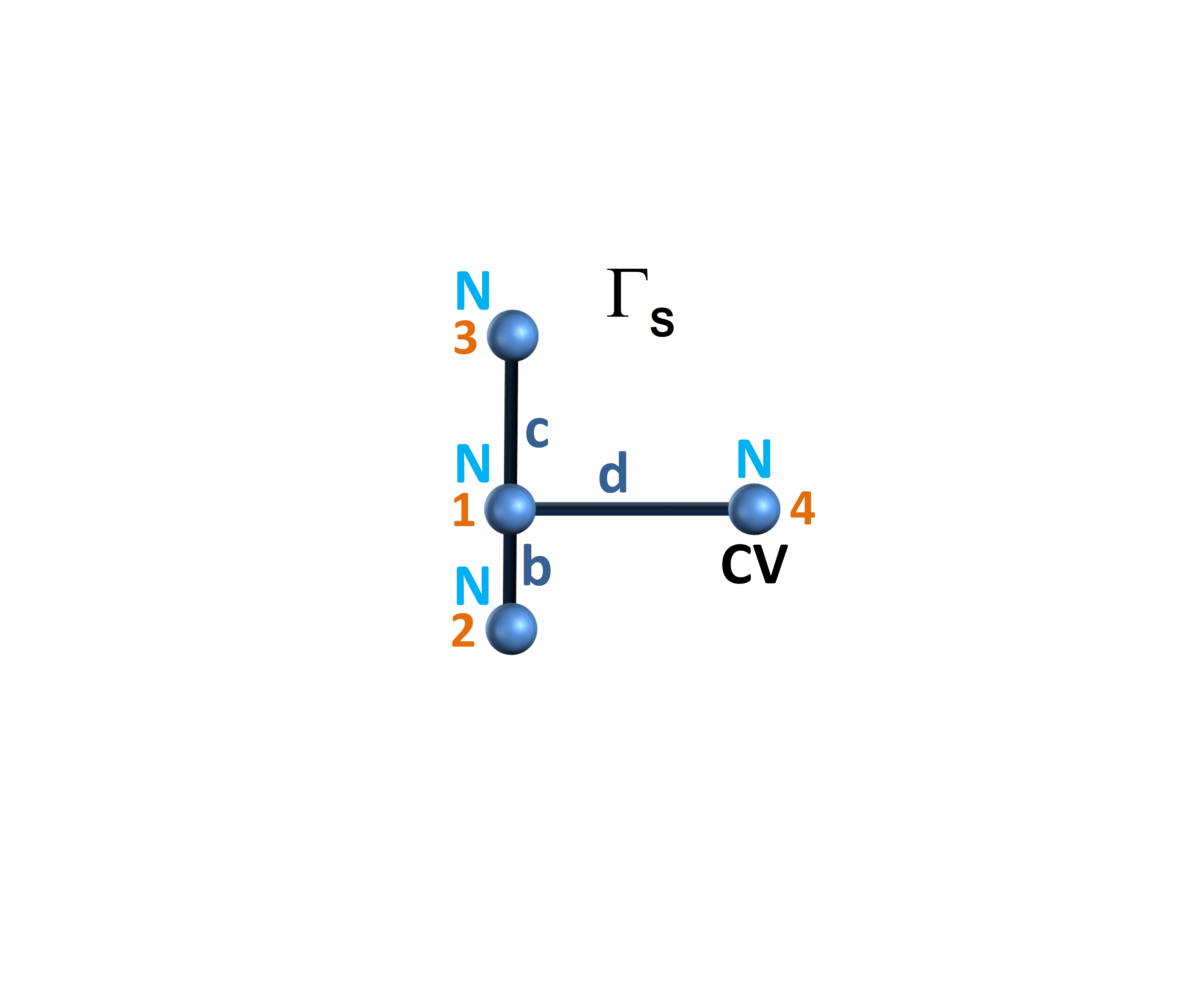}
		\caption{
			A star graph $\Gamma_{\rm S}$ with three edges $b$, $c$, and $d$ and Neumann boundary conditions at the vertices, which are indicated by blue capital letters N. The contact vertex is marked by the symbol CV. }
		\label{FigS2}
	\end{center}
\end{figure}

	A star graph $\Gamma_{\rm S}$ with three edges of lengths $b$, $c$, and $d$, and  
	standard vertex conditions is shown in Fig.~\ref{FigS2}.
	The contact vertex is denoted by CV.
	
	Assume that all edges are parametrized in the direction towards the contact vertex.
	Then on the pendant edges the eigenfunction is given by
	$$ u (x) = a_j \cos kx, $$
	which ensures that standard ({\it i.e.}, Neumann) conditions are satisfied at the degree one vertices.
	Hence the M-function for pendant vertices is
	$$ M_{{\alpha}^{\rm N}} (k) = \frac{\partial u (x)}{u(x)} = - \frac{u'(x)}{u(x)}  = k \tan k \alpha, \quad \alpha = b,c, $$
	where we took into account that the oriented derivatives should be taken in the direction inside the graph.
	
	Hence the M-function for the graph formed by the two left edges and the central vertex as contact point is
	$$ M_{{\rm (b+c)}^{\rm N}} (k) = k \tan k b + k \tan k c. $$
	The eigenfunction on the third edge should satisfy  the energy dependent condition
	$$ \frac{u'(0)}{u(0)} = - M_{{\rm (b+c)}^{\rm N}} (k) $$
	to ensure that standard conditions at the central vertex are satisfied.
	Such function is given by
	$$ u(x) = \cos kx - k (\tan kc + \tan kb) \frac{\sin kx}{k} . $$

The M-function for a star graph $ M_{\rm S}(k)$ is obtained by computing the normal derivative of $u(x)$ at $x=d$
	\begin{equation}
 M_{\rm S}(k) = \frac{\partial u (d)}{u(d)}  = - \frac{u'(d)}{u(d)} =
 k \frac{\tan kb + \tan kc + \tan kd}{1 - \tan kd (\tan kb + \tan kc)} . 
	\tag{S3}
	\label{S3}
	\end{equation}

\newpage

